# INTERACTION, CHANGE, AND WHOLENESS OF MATERIAL THINGS

## Oleg G. Semyonov

*State University of New York at Stony Brook, 214 Old Chemistry, Stony Brook, NY 11794, USA*
osemyonov@ece.sunysb.edu

Abstract

Relational holism is an inalienable feature of the material world both on microscopic and macroscopic levels of matter composition. The driving force of change is interaction of things. A thing is acknowledged as 'this' with its specific properties through specificity of changes of other things occurring to them in interaction with this thing. In a system of interacting things, all the constituents co-influence and co-change each other; the changes of interacting individuals become interdependent and interdependency of their changes results in interrelation of their properties. Interrelation culminates in formation of ensembles – material wholes of correlated things. Wholeness is the consequence of interdependency of changes of parts in their together-being. Every whole emerges through mutual correlation and togetherness of interacting individuals; in an ensemble, the mode of being of a particular component depends on the modes of being of all other components and vice versa. Every ensemble attains its wholeness and becomes a physical object through togetherness of components coexisting in the form of a collective being.

## 1. INTRODUCTION

The worldviews of ancient thinkers (Aristotle, Ptolemeus) and eighteenth-century scientists (Newton, Laplace) mirrored the levels of scientific knowledge of their epochs: the metaphors of the world were either sophisticated systems of interconnected gears or mechanical clocks. Owing to the prominent advance of general relativity and cosmology during the second half of 20th century,(Rindler 2006, Roos 2003) a significant progress was achieved in understanding of our Universe as an evolving system of correlated macroscopic objects. In its turn, the quantum-mechanical Standard Model (Nachtmann 1990) became a fairly elegant description of matter on a microscopic scale ~ $10^{-15}$ m. String theory (Smolin 2007), quantum gravity (Smolin 2002), and theory of noncommutative spaces (Heller 2000) will, hopefully, culminate in enrooting the structure of matter even below Plank length (~$10^{-35}$ m) with potential unification of micro-, macro-, and mega-worlds in one Grand Unification theory. The impressive progress of our knowledge about the material world owes its success to universal acceptance of physics as a science of interaction, where interaction is understood as an inherent ability of material things to distantly influence each other by means of specific entities, namely, physical fields continuously spreading over geometric space between discrete things. These fields are thought to be the 'agents' of influence, who are actually responsible for all changes of things, for to be influenced is to be changed. No ontological rationale of the concept of physical field has ever been suggested, nevertheless, it became a cornerstone of modern physics because it explains distant influence and makes performable numerical evaluation of action.

It is intuitively clear that wholeness of material things is somehow linked with relations (read interaction) between their constituents. Methodological holism, which encompasses both relational holism and structural realism, declares that in amalgamation of parts a whole attains properties that can not be derived from the properties of its own components, while reductionism implies the principal possibility to deduce all the properties of a whole on the condition that all 'intrinsic' properties of parts and all their relations are taken into account. Relational holism (Teller 1986) claims that "there are non-supervening relations… that is, relations that do not supervene on the nonrelational properties of the relata." In their turn, reductionists (e.g., Weinberg 1992) insist, for example, that thermodynamics can be explained in terms of particles and forces, which would hardly be possible if the thermodynamic laws were autonomous; the notorious uncertainties of mechanical parameters of individuals in the multi-component systems are thought to be linked with our failure to take into account all relations of parts including relations external to the systems.

The philosophical debates between those who support atomism and those who defend methodological holism are mainly focused on attainability of at least some properties of a whole which are irreducible to the properties of its own parts or, in other words, on the emergent nature of the properties of wholes; ontological emergence was actively discussed in opposition to atomism and supervenience.(Andersen at al. 2000, Lewis 1986, Teller 1986) Both, relational holism and structural realism pretend to be ontological explanations of wholeness despite their disagreement, in particular, in their approach to intrinsicness: relational holism admits existence of intrinsic (non-relational) properties of things while structural realism claims that there are no fundamental intrinsic properties underlying the relations:

> It is quite consistent to suppose that two distinct individuals, each having a non-relational property, should also stand in some inherent relation to each other (Teller 1986).

> If it is true that our basic physical theories give us knowledge only of the relations in which physical objects stand, the metaphysics of intrinsic properties is in trouble: metaphysics has it that the world consists of objects that are characterized by intrinsic properties each. On epistemological reflection, however, we have to concede that we do not have access to these properties insofar as they are intrinsic (Esfeld & Lam 2005).

It should be noted in this connection, that explicit division of things' properties into relational and non-relational is, in many aspects, subjective despite its apparent obviousness. Absolute 'intrinsicness' (Yablo 1999), i.e. existence of properties without others, beyond perception or detection, and independently on accompaniment or loneliness (Langton & Lewis 1998), is an unproven speculation because to be intrinsic means to be non-relational and, therefore, undetectable and unknowledgeable. There is no factual knowledge about the material things obtained other than through perception or detection.(Esfeld & Lam 2005) The properties, commonly referred as non-relational, come to 'obviousness' through relations. We ascribe the non-relational properties to the material objects (more exactly, we consider them intrinsic and non-relational) inferred from their perceived, detected, and measured relational properties; intrinsicness, as such, dissolves.

Emergence of properties of material wholes is believed to be a real feature of matter in the domain of quantum mechanics (Teller 1986) and in biology (Andersen at al. 2000, Campbell 1974), where irreducibility of properties of complex dynamic objects to properties of their components, taken separately, is more or less self-evident. The non-biological macroscopic objects are mainly regarded as the statistical systems of the individuals obeying Hamiltonian mechanics in its canonical form. It is not of surprise that reducibility of these systems seemed to be proven because Hamiltonian mechanics produces, in principle, the exact solutions for any number of interacting things. A strange gap of reducibility and atomism is set in between the holistic worlds of quantum-mechanical micro-objects and macroscopic bio-organisms.

In this paper, a role of methodological holism in micro, macro, and mega-worlds is discussed. Material wholes are considered as the collective beings of the interrelated parts. Interrelation and formation of material wholes are regarded in the context of distant mutual action of things. Though some philosophers considered the concept of distant action nonsensical (Newton 1999, Leibnitz 1890, Bedford 1893, Turnbull 1959) (see also a remark of Teller 1986, 74), this concept can be fairy logical, if we either introduce a specific material entity (physical field) as a carrier of action related to every point of space [Teller 1986, p. 76] or dare to dismiss our old view on space as something absolute, uninfluenced, and unchanging in favor of envisioning it as a kind of material entity capable of alteration in the presence of the discrete things (see details in Section 3.1). Ontological significance of the notion of interaction in the material world, which is envisioned as a hierarchy of wholes, is discussed with emphasis of a link between change, emergence, and wholeness.

## 2. CHANGEABILITY AND WHOLENESS (GENERAL REMARKS)

Changeability and wholeness – the capacities of material things at first sight unrelated to each other – have been debated since the very beginning of cognitive contemplation of Nature.(Aristotle 1941, Bohm 1980, Hegel 1892 and 1969, Heraclitus 2001, Meirav 2003, Mortensen 2002, Popper 1998) "All change is change of something" (Popper 1998) and, obviously, there must be something that causes the change. The world consists of nothing else but individual material things and it is logical to assume that the cause of change lies in the things themselves. No material thing stands alone. A sole (isolated) thing is our mental abstraction. Others are always and everywhere in the real world with respect to every particular 'this'. Existence of something can be noticed by others and acknowledged by a curios mind on two conditions: first, this something has an ability of sending a signal to others, 'Hey, I am here', and second, other things are able to 'sense' this particular thing, i.e. to receive and to discriminate the signal. To receive a signal means to undergo some change of the mode of being which means alteration of the properties. To receive is to react and to react is to alter. Every change is a provoked response of a thing to a particular signal sent by a particular another. It stands witness to reality of the incoming signal and, therefore, to existence of the signal-sending thing because it would be hardly sensible to think of the signal that causes reaction and change of the material things without something material behind it. The spooky arguments in favor of 'uncaused change' (Mortensen 2002) with reference to radioactive decay demonstrate simply the common misunderstanding of the processes of spontaneous emission (see details at the end of Section 3.2).

The change of an observed thing is the only means for discerning a particular signal and, ultimately, for acknowledging and evaluating a particular property of another displayed by this signal. Actually, we ascribe a particular property to a particular signaling thing after observing a particular change of a receiving thing, assuming that this particular change mirrors the particular property of the signaling thing. Selfness of a thing is attributed to it through reactions of others, meaning that other things specifically change their properties in comparison with their pristine properties that would be without this thing. As the existing things, all others also send their backward signals to the first signaling thing, and, which is of extreme importance, their reciprocal signals differ from the signals that would be without this thing because others have been already changed due to their reaction to this thing. Reaction of others to this thing, i.e. their change, leads inevitably to some change of signals sent by each of them outward. This alteration of signals means that their action on all others, including the first influencing thing, is altered in comparison with their influence that would be without the first influencing thing. All the material things around a particular thing undergo the mutually conditional changes in their responses to the conditionally altering influence of this thing while this thing undergoes the conditional change in its response to the conditionally altering influence of others.

The conventional vision is that there are the properties inherent in a particular thing independently of others and that these 'intrinsic' properties are uninfluenced and, therefore, unchanging. In fact, we obtain information about all the properties of a thing from observations of other things: first, a particular change of another is picked out and judged as its response to something influencing, then we come to conclusion of the presence of the signal (action) sent by this something to the observed thing to cause its change, and finally we ascribe a specific property to this something as our intelligent guess based on our observation of the specific change of another. The influencing thing is postulated to be "the bearer of the properties" (Heidegger 1962, 1967) as if the properties are frozen in it irrelevantly of others and relations with them. It should be remembered, however, that every particular property of every particular thing is its "being-for-another" (Hegel 1892 and 1969) and has no sense without others and relations with them. "Being is always the being of a being"(Heidegger 1967) which can not be acknowledged and the properties of which can not be determined without observing the responses of others to its being. Being is always the being of a being amid other beings. "Somewhat is by its quality, - firstly finite, - secondly alterable" (Hegel 1892) "and in this change the others are involved" (Aristotle 1941). The controversy of the notion of intrinsicness was debated in numerous publications. (Langton & Lewis 1998, Esfeld & Lam 2005)

The only way-of-being of an individual thing in the material reality is its interaction with other individuals. Every material thing is the source of influence on others and the cause of their change. Reciprocally, every material thing is the object of influence by others and undergoes changes caused by others. In other words, all coexisting things permanently interact with each other meaning that their changes become interdependent through the conditionally altering influences on each other. Existence of an individual thing is always its coexistence with others in the mode of mutual influence and interrelated change. Being of things is their being in the mode of co-influence and co-change.

The prominent capacity of all material things is their ability to form material wholes consisting of components. "Almost anything we care to think about may be considered as a whole and as having parts, or as being itself a part of some greater whole." (Meirav 2003) All physical objects are finite: they occupy finite volumes in space and consist of finite number of constituents. A composite object is to be understood not as simple composition of parts but as unification of interacting components, which are finite physical objects, themselves, in wholes. Interacting elementary particles form nuclei, interacting nuclei and electrons form atoms, interacting atoms form molecules, interacting molecules form macroscopic physical bodies and substances, interacting physical bodies form celestial objects, and, eventually, interacting celestial objects form our Universe as a structural superwhole. It is more or less obvious that the randomly sampled material things do not necessary compose a functional or/and a structural whole even if they belong to a class or a set of elements selected by a certain criterion. An arbitrary selection of live cells does not necessary produce an organism, a mixture of atoms or molecules of various elements does not necessary becomes a chemical compound, and patches of gas and dust in space do not necessary form a solar system. To compose a whole, its parts must be in some kind of bounding influence on each other. This 'bounding influence' is the key phrase for understanding interrelation between wholes and parts. To form something whole, all parts must be tied one way or another not in a sense of their belonging to a selection by their individual 'intrinsic' characteristics and not because of their apparent similarity, but as their coexistence in the mode of mutual link, mutual influence on each other, and mutual constraint of individuality. The parts loose their independency and isolation from each other and obey the 'rules of game' specific for the whole they form. This obedience means that all parts are united in something more complex than every lone individual or a set of independent individuals. It can be said that the parts are entangled in a whole. Every whole is also something: an atomic nucleus is formed by elementary particles bounded by nuclear forces that confine them in a finite volume of space to give birth to a sizable entity; a molecule is formed by atoms tied by electromagnetic forces that hold them together within a structural object; a physical body is formed by molecules mutually bounded by electromagnetic forces, etc. The actuality that dictates unification of material things in wholes is interaction. It is the cause of their mutual change and the reason for their interdependency culminating in their togetherness which is interpreted as wholeness.

## 3. TOGETHERNESS THROUGH CHANGE

**3.1 To be is to interact**

The concept of interaction was born from our observations of material things (hereafter, things) around us to denote our acknowledgement of their ability to distantly act on other things and to undergo the balanced actions of others. In mechanics, this phenomenon is called the $1^{st}$ Newton's law: action is equal to counteraction. The balanced mutual influence of things and the balanced change they lay on each other does not mean that all the enumerated changes of their properties are necessarily equal. It means rather that every isolated (not subjected to external influence) system of interacting thing keeps the averaged properties of its constituents such as their average mechanical momentum and total relative energy unchanging while the instant properties of each of them are permanently altering because of their mutual influence.

The conundrum of distant action of the discrete material things on each other was always a stumbling-block in metaphysics. The commonly accepted metaphor of space as an absolute and unchanging void requires a carrier of action capable of spreading across and propagating through space, i.e. a kind of continuous material substance overlapping space (immaterial carrier has to be ruled out due to inability to produce action on anything material). This carrier is supposed to be 'emanated' from a discrete thing in order to act on other distant things and to force them to specifically change their physical properties. The vision of space, persisted over centuries, was a continuous manifold of geometric points of zero size (with the inherent logical controversy between the zero-sized points and the extension produced by them) which exists independently of matter and, because of that, stays uninfluenced and unchanging, ergo immaterial and eternal. According to this metaphor, the discrete things are embedded in this absolute geometric space as spatially separated entities, each having a finite geometric size. Spatial separation and discreteness presuppose occupancy of a finite volume non-overlapping the volumes occupied by other things, which means that there is some 'free of matter' space between the things. Obviously, the concept of free-of-matter space is inherently inconsistent and this inconsistency was always disturbing. The very fact of influence of things on each other implies that space between them can not be a void or an absolute (unchanging) entity because nothing is able to propagate through nothing or through unchanging something and no influence can be actualized. It must be something changeable to convey the influence, i.e. a kind of material carrier of action subjected to change in the presence of the discrete things and conveying, by means of its own change, the changing influence on other things, "since neither has any share of nothingness" (Parmenides 1986) and "either, then, nothing has a natural locomotion, or else there is no void" (Aristotle 1941). Newton's ether – a continuum of mutually contacting absolutely rigid material points of zero size and zero mass, uniformly filling space between the discrete things, – was a compromise between the divine space and the material reality: some materiality was attributed to space while its changelessness was left intact. Something can not, however, be a little bit material but otherwise immaterial. It is either material or not. If it is material, it can not be uninfluenced and unchangeable; if it is immaterial, it can not carry material properties such as rigidity postulated by Newton, even if this property was declared 'absolute'. Absolute rigidity means nothing else but mechanical changelessness and thus inability to transfer mechanical action associated with alteration of density and pressure.

Three fundamental entities constituted the Newtonian world: absolute geometric space, discrete matter, and continuous unchanging matter possessing absolute rigidity (ether). Newtonian ether was ruled out by the Leibnitz's idea of continuity of things. He considered space between the discrete things as a mere property of things extended beyond their visible boundaries and as their distant 'activity': "I agree that naturally every body is extended, and that there is no extension without body… Besides extension, there must be a subject which is extended, that is, a substance to which it belongs to be repeated or continued" [Leibnitz 1890]. This idea found a new twist when Maxwell learned that equations for electromagnetic waves in space require no material point or body; mathematically, their behavior could be completely determined by their own contiguous processes (if we forget for a moment that an electromagnetic wave must be excited, first of all, by an oscillating (moving) electric charge associated with discrete matter). A concept of physical field emerged – a material entity which is linked somehow with a discrete thing and which continuously fills geometric space between the discrete things to actually produce the distant action. The concept was successfully applied to all known types of action: nuclear, electromagnetic, and gravitational. Its conquest in physics was so impressive that many scientists, including Einstein obsessed with the idea of unification, postulated these continuous fields as the only physical reality. Discreteness was ruled out and discrete material objects were considered simply as local regions in space of relatively high 'strength' of field, putting aside the fundamental question about the cause of such local elevations of fields and their stability. The content of the world was reduced to two fundamental constituents: physical field (continuous matter) and absolute geometric space.

It is obvious that space as a receptacle for discrete matter performs twofold function: firstly, it separates things into different 'here' and 'there', granting them an ability to exist as separate geometrical objects characterized by their sizes, relative distance, and relative motion (change of relative distance), and secondly it allows their distant influence on each other thus making possible alteration of their properties including change of their relative mechanical motion. For an observer, space between a couple of things transfers influence of one thing on another for recognition, response, and change. Since space was postulated absolute, i.e. unchanging and therefore unable to transfer influence, it is not of surprise that a carrier of influence had to be added to explain the distant action. To transfer influence, this carrier must be 'emanated' from an influencing thing and carry information about its properties to others for recognition and response when it reaches them. The insight of common physics on the nature of this carrier is the mentioned concept of physical field, that is, a continuous material entity spreading over the undisturbed absolute geometric space between the discrete things. This view, however, returns us back to Leibnitz: every physical field can be considered as an extended part protruded from a discrete thing (magnetic field is an exception, however it is not a primary field but a progeny of an altering electric field) and nothing can prevent us to regard the physical fields as mere attributes of things. Despite its mathematical beauty and universal acceptance in science, the field concept suffers from one yawning metaphysical controversy: material fields exist in and spread over absolute, that is, immaterial geometric space. Such a metaphor seems to be just another desperate effort to sew together matter and 'divine' space. Absolute geometric space is our mathematical idealization – a mental model emerged from our perception of arrangement and movement of things around us with their visible sizes and separation from each other. There is a growing conviction in modern science that space itself is to be considered as a kind of material entity able to interact with conventional discrete matter and, because of that, subjected to change, i.e. it is something possessing an ability to be distorted in the presence of the discrete things. The metaphor of space as a material substance does not need fields for transferring actions. If the field concept is dismissed, the content of the world is reduced to one universal entity, namely, matter which can exist in two manifestations: discrete matter (things) and continuous matter, perceived by us as space (the definition 'continuous' should not be understood as infinite division of a volume into smaller and smaller parts; modern physics considers space as a sort of quantum substance consisting of discrete modicums of the order of Plank length ~ $10^{-35}$ m indivisible into smaller parts (Smolin 2007 and 2002, Heller 2000)). Space is influenced by discrete matter and undergoes change, and it acts on discrete things to cause their change. For example, the theory of general relativity treats space as a changeable object. Space becomes distorted (bended) around a discrete thing that possesses a finite gravitational charge (mass) meaning that every massive thing changes metrics of space. In contrast to the concept of physical field in absolute space with flat metrics, the curvature of space produced by a massive thing is actually responsible for the gravitational action on other things, which is, as it was proven by Einstein, completely equivalent to the description of action in terms of gravitational field in Euclidian space.(Hartle 2002) The bended space around a massive thing produces mechanical action on other massive things; those, in their turn, bend space around each of them to mechanically influence other discrete things, and so on. Extending this vision to other known types of action, it can be argued that space as a material entity responds to conventional discrete matter and becomes tensioned when a discrete material thing is present: massive things cause space bending, electrically charged things cause electromagnetic polarization of space[1](Moriyasu 1983), and things, possessing baryonic charge, cause specific tension of space in their close vicinity interpreted as nuclear field.(Blokhintsev 1973) The phenomenon of influence of one thing on another can be understood as a tendency of disturbed space to be completely symmetric, i.e. as its propensity to lessen the asymmetry of strain, if space is tensioned by two or more neighboring discrete things as shown schematically in Fig. 1. Without its counterparty, every massive thing produces symmetrically curved space around it (Fig.1a) spreading to infinity with gradually diminishing curvature. The backward tensile mechanical action of space on this thing is isotropic and results in zero total mechanical force applied to a sole thing and, therefore, in indifference of its spatial position or motion. When another massive thing is present (Fig.1b), the curvature of space between them differs from the curvature produced by a lone thing. Space around each thing is now bended asymmetrically resulting in asymmetric tension and therefore in asymmetric tensile force applied to each thing from this asymmetrically strained space. In the case of gravitational bending, the maximum angular symmetry of space curvature can be achieved when both things collide together. Thus, a couple of initially immobile and distantly separated massive things are pushed to each other by the tensile force of asymmetric space which results in their movement toward each other. This push is interpreted as an attractive gravitational force produced by gravitational field. The force of action diminishes, when the distance between the things increases; at longer distances, each thing is positioned in the regions of lesser curvature produced by its counterparty and, therefore, in the regions of lower induced asymmetry of tensioned space, which means that the mechanical force acting on each of them is weaker.

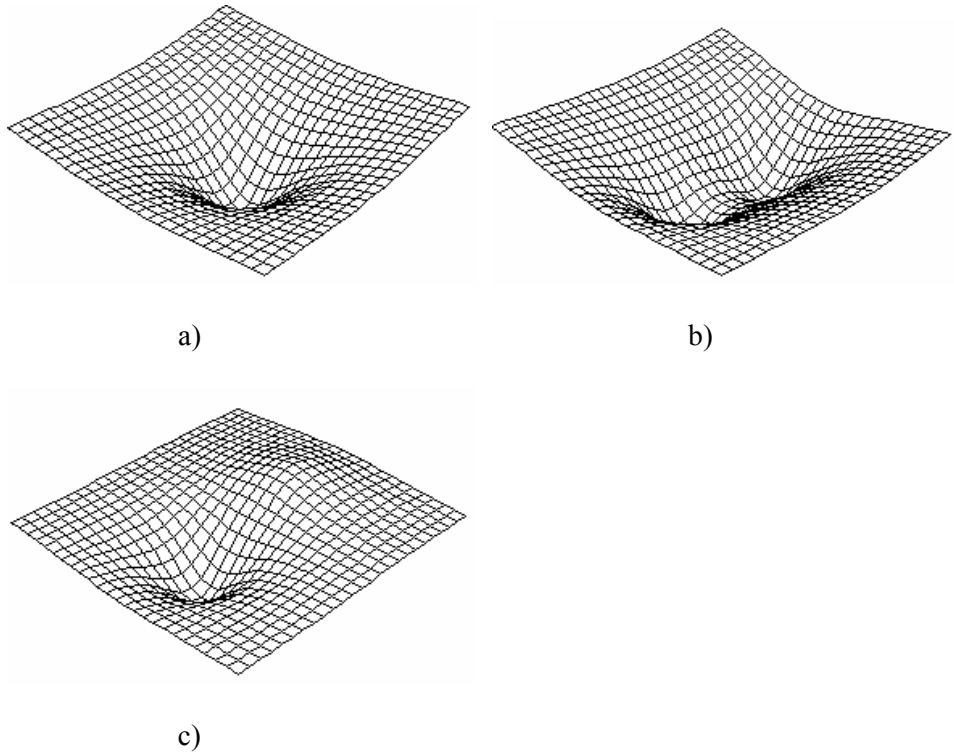

Fig. 1. Schematic 2D representation of space bending caused by a) a sole massive object, b) two neighboring massive objects, and c) two antigravitating objects (one of them has a hypothetical negative mass).

It is reasonable to assume that the integral tensile force applied to a spherical surface enveloping a sole thing in space is the same irrespective of a radius R of this sphere (otherwise it would be no equilibrium state of space). The distortion of space per unit area and, therefore, the tensile force applied to another thing at a distance R from this thing diminish as $1/R^2$ on the condition that space curvature is not extremely high, which means that the radius of curvature is large in comparison with the distance between the things. On this condition, space can be considered nearly flat and the total area S of a sphere of radius R is proportional to $R^2$ (an exception is space in close vicinity of a black hole where space metrics is highly distorted), meaning that the force applied to a remote thing with its cross-section A, is proportional to $A/S = A/4\pi R^2$. This is completely equivalent to the well-known Gauss's theorem for the physical fields in a flat (Euclidian) 3D space, yielding the same $1/R^2$ dependence of a force acting between two interacting things. [2]

A hypothetical negative mass would produce an oppositely bended space which can be imaged as a hill in Fig.1 instead of a pit. Two negatively massed things would be attracted to each other because the patch of space between them would be less strained in comparison with other directions. However, interaction of a thing of negative mass with a positively massed thing should be repulsive because space between them would be bended steeper in comparison with its curvature in other directions: the zero curvature of space is achieved at a finite distance between them while at infinity in all other directions, as shown schematically in Fig.1c.

Likewise, a sole electrically charged thing induces some symmetric polarization of space, which can be envisioned as the preferential virtual separation of the paired (annihilated) electron/positron pairs in Dirac vacuum (Moriyasu 1983) accompanied by the directional priority of their vibrations. The virtual photons in space, appearing for an instant and then disappearing as a result of vibration of these pairs, will be polarized with the electric components of electromagnetic field collinear to the line connecting a given point in space and the electrically charged thing. If another electrically charged thing is present, a patch of space between them becomes more polarized (more strained out) to produce attraction when their charges are opposite, or less polarized with respect to other directions (with zero polarization of space at some point between them) to beget repulsion when their charges are of the same sign. Nuclear interaction of baryonic particles can also be considered as the reaction of space on quantum level in their close vicinity, resulting in the tensile strain accompanied by the preferential polarization of the appearing virtual matter/antimatter pairs and in asymmetry of gluon clouds (distribution of probability of quarks to be at a particular point of space at a particular moment in the vicinity of a particle) around each of the particles when they are in proximity to each other. The space tension induced by quarks or elementary particles (nuclear field associated with it) diminishes faster than $1/R^2$ due to essentially quantum nature of particle/vacuum interaction; nuclear interaction shows itself mainly at short distances $\sim 10^{-15}$ m (Blokhintsev 1973).

When a region of tensioned space around an influencing thing overlaps a region of tensioned space created by another thing as shown in Fig.1b, the space tension around an influenced thing looses its original symmetry because the patches of space closer to the influencing thing become stronger disturbed (because of the $1/R^2$ dependence) in addition to the distribution of space tension created by the influenced thing while the patches of space behind it are additionally disturbed to lesser extent. As a result, a region of asymmetrically distorted space is formed around the influenced thing. It forces this thing to move along the line of maximum asymmetry of the disturbance (a geometric line connecting the influencing and the influenced things) in the direction of the lowest gradient of space tension. Reciprocally, the region of tensioned space created by another thing overlaps the region of tensioned space around the first thing and produces some asymmetry of space tension around the first thing. It forces the first thing to move toward or from another thing along a line connecting both interacting things. A type (attractive or repulsive), as well as strength of the acting force depend on a gradient of space distortion. If this gradient is smaller between the things then behind each of them, the force is attractive and the things tend to move to each other (Fig. 1b). If the gradient of space tension is higher between the things and the state of undisturbed space is reached at a finite distance at some point between the things while at infinity in all other directions, the force is repulsive and the things tend to move away from each other (Fig. 1c). The degree of disturbance of space produced by a particular discrete thing is characterized by a specific parameter called charge. In the case of gravitational bending, the charge is mass and the larger is mass, the higher is the space curvature around this mass at a particular distance from it. In the case of electricity, the charge is electric charge and the larger is electric charge, the higher is electromagnetic polarization of the annihilated electron/positron pairs around it, the higher is the tension of space at a particular distance, and the stronger is its influence imposed on other electric charges. The repulsive or attractive forces acting between the discrete things cause their relative movement resulting in alteration of distance between them. Alteration of distance leads to alteration of their action on each other due to changing asymmetry of tensioned space around each of them. Changed asymmetry of space tension results in modified reaction of things (slower or faster acceleration), which leads afresh to some change of mutual influence due to altered space asymmetry around each of them, and so on; the things interact and display their mutually coordinated motion.

The concept of materiality of space is, apparently, the only consistent explanation of distant action. Moreover, it makes understandable the principal possibility of unification of different types of actions (fields in the field theory) because every sort of action is linked with space tensions. The unified field theory unifies nuclear fields (strong and weak) and electromagnetic field in one universal field. Gravitation has eluded so far unification with other types of fields and, perhaps, this fact is the consequence of principally different reaction of space to mass in comparison with its reaction to electric and baryonic charges: nuclear and electromagnetic interactions are associated with polarization of space (vacuum), i.e. with 'longitudinal' strain of space which does not influence space metrics, while mass causes space bending, i.e. 'transverse' tension that alters space metrics.

Materiality of space is one of the basic premises of modern quantum theories. From the quantum mechanical point of view, discrete matter can be envisioned as consisting of elementary particles (ensembles of quarks) that come to existence as a result of excitation of matter from the lowest possible quantum energy level (vacuum state) to some higher allowable potential energy level. Space is the perceived form of quantum vacuum in the presence of the discrete things while quantum vacuum itself is the degenerate form of matter on the lowest quantum energy level; physical vacuum can be considered as a quantum ensemble of the annihilated matter/antimatter pairs of particles.(Moriyasu 1983) Annihilation of matter and antimatter is transition of the material modicums to the lowest quantum energy level of matter existence accompanied by some energy release due to changing potential energy of interaction of components (e.g. quarks) in these particles. This energy release is perceived in the real space as emission of electromagnetic waves (oscillatory change of space polarization), i.e. photons in quantum-mechanical language, or as generation of other bosons. The reverse process of birth of pairs of real particles from vacuum, for example, the birth of an electron-positron pair by a gamma photon, is the transition from the lowest energy level to a higher level (from the vacuum state to the particles-in-space state) accompanied by energy absorption. Space and discrete matter are actually two different states of matter existence: a) physical vacuum of annihilated discrete matter on the lowest possible quantum level and b) discrete objects of the perceived reality when the underlying vacuum entities are excited to some higher quantum energy levels and become the real particles (ensembles of quarks) separated by space. The perceived space is the way of existence of the discrete things with their locality and finite sizes on the background of quantum vacuum. Space acquires its extension in the presence of the discrete things and has no sense without them. Distance (enumerated extension) measures separation between the discrete things and it is nonsensical without discrete matter. The finite size of a thing as an ensemble of interacting components and spatial separation of things (ensembles) mean that their interaction is always distant. For example, mechanical collision of two solid objects is not their merging at a point of contact, but their rapprochement to a distance of action of intermolecular forces approximately equal to the spatial separation of molecules or atoms in them.

It is important to mention that flatness attributed to absolute space (without matter) demonstrates simply our dogmatic vision. It follows from nowhere and exhibits our belief in infallibility of the fathers of science whose vision was based on mere sensual perception. A mixture (more exactly a unity) of space and discrete matter in our Universe is virtually always non-flat with the exception of one specific case, namely, when the sum of average densities of massive matter $\rho_M$, radiation (photons) $\rho_R$, and mass-density $\rho_V = E_V/c^2$ of so-called vacuum energy $E_V$ (dark energy), where c is the speed of light, is exactly equal to critical density $\rho_c = 3H^2/(8\pi G)$, where $H$ is the Hubble constant at a given cosmological time of the Universe's evolution (e.g., at a moment of existence of an observer, i.e. us) and $G$ is the gravitational constant.(Roos 2003) Friedman-Robertson-Walker (FRW) metrics of space is described by a tensor that exhibits in complete generality the manner in which space is curved. A single parameter k is introduced to distinct spaces of negative curvature (k = -1 when $\rho_M + \rho_R + \rho_V < \rho_c$, for example, space without massive matter), positive curvature (k = 1 when $\rho_M + \rho_R + \rho_V > \rho_c$), and flat (k = 0 when $\rho_M + \rho_R + \rho_V = \rho_c$). Astrophysical observations are consistent with $\rho_M + \rho_R \approx 0.3\ \rho_c$ (including the so-called dark matter) and $\rho_V \approx 0.7\ \rho_c$ at the current epoch with an accuracy of approximately 2 to 3%, which means that space in our Universe is, on average, close to flat. However, it doesn't mean that it is exactly flat because even a miniscule deviation from the exact equality would yield space of either positive curvature (spatially finite and gravitationally closed universe) or negative curvature (spatially infinite and gravitationally opened universe). In this sense, the absence of conventional matter ($\rho_M + \rho_R = 0$) would not result in flatness of space. On the contrary, this space would be negatively curved except we postulate again $\rho_V = \rho_c$, which does not seem to be the case. Space possesses mass (energy) just as conventional matter!

As it has been mentioned, a thing is perceived, detected, noticed, and observed through its action (mechanical, chemical, electromagnetic, gravitational, etc.) on the receptors of our senses or on other things observed by us. Our receptors, detectors, measuring tools, i.e. the things with their properties known to us from earlier experience of interaction with them, or other physical things 'sense' a concrete thing, response to its existence, and discern its properties solely through the changes of their own properties because every change of their properties induced by an influencing thing is nothing else but their specific response to the changed spatial environment, namely, to the changed asymmetry of tensioned space around them. A particular change of another is judged as its reaction to the 'bearer' of the specific properties because of the specificity of the reaction. Availability of things with their anticipated reaction to a particular influence is prerequisite for acknowledgment of a particular thing and for studying it.

In general, interaction is the necessary aspect of thing's existence. Without ability to influence others, a thing can not be perceived, recognized, and acknowledged; without its ability to be influenced, it can not be changed and, which is most important from the practical point of view, it can not be handled. An inactive, i.e. non-interacting, thing does not exist for others. It is nonexistent for the world of things and for us, too, because, without reactions of other things, no change of others in their response to its existence can happen. Without change of others, this thing can not be noticed by others and acknowledged by an observer as existing something. The very possibility to be detected (perceived), i.e. the potentiality of a thing to cause some reaction of another to its own beingness, predetermines its existence as 'a thing'. Only about an interacting thing one can say 'it is'. Nothing can be said about an isolated and, therefore, non-interacting thing. We can prove its existence by no means, since if it does not exist for others and for the material world in general, it is the same as it does not exist at all. [3] If, on the other hand, there are no others and no environment, a thing can not realize itself as a being, because its quality, if assumed, can not be presented to anything when nothing is able to perceive it and to react to the existence of this hypothetic bearer of quality. Others and relations with them are the necessary elements of existence of every 'this' as an acknowledgeable entity. 'The determinateness through which one thing is this thing only, lies solely in its properties', but 'property exists only as a mode of relationship between things'.(Hegel 1969) Others and relations with them fashion existentiality of things as the bearers of the properties. Determinateness of properties emerges through specificity of relations (interaction); the non-relational properties would be rather the characteristics of the non-interacting abstract beings. We postulate existence of non-relational properties beyond relations referring to the principle of causality because it seems logical to explain the change of an observed thing as its reaction to something that possesses its pristine, unchanging, and independent properties somehow demonstrated by the observed influence. Pushing a thing by another thing and watching its acceleration (change of relative velocity), we produce action (mechanical push or force) $F$ and watch reaction (acceleration) $a$. A ratio of enumerated action $F$ and reaction $a$ gives us a proportionality coefficient $m = F/a$ between the applied force and acceleration, called 'mass'. We do not feel or observe mass; we observe change of movement and, being perplexed with the fact that our push does not produce a momentary change of state of movement but some graduate alteration, we ascribe a property of inertia to this thing and characterize it by a mass $m$ as existing in this thing beyond observation without questioning themselves where it actually came from. When we observe a stick having a length $x$, we either sense its image created by the eye lens on the retina (light reflected by the stick and imaged on the retina) or feel it with a hand. How do we know that a thing has a length x or a mass m? Solely by entering in interaction with this thing using our senses, detectors, or measuring tools and referring it to another thing of known length or mass taken as a unit, for example, to the standard meter or the standard kilogram stored in National Bureau of Metrology, Paris. Without another material thing, even if another is our body, for interaction and comparison, we would not be able to ascribe a length to a stick or a mass to a physical body; actually, even the very idea of length or mass would not appear, if there were no others for interacting and scaling. We have to sense a stick to become aware of its length and to sense a body to evaluate its mass, in other words, we have to interact, i.e. to be in a sort of action-counteraction relation. [4] Counteraction means inevitable influence of a detecting tool (influenced thing) on a detected (influencing) thing and unavoidable change of the properties of this influencing (measured) thing, too. It means that the properties of measured things are not pristine and unchanging, so it is not obvious do we measure an explicit enumerated property of this thing and can we measure anything explicitly at all. This controversy, actively debated in the second half of XX century, is linked to the problem of accuracy of measurements, which, in its turn, is connected with the mental omission of the changes of measured objects caused by our measuring tools. The measurability problem is the most vivid for the microscopic quantum-mechanical objects where the influence of our detectors on the measured objects can not be taken negligible (Bohr 1934, Einstein 1949, Bohm 1980, Healey 1989).

As for the problem of intrinsicness, the logical solution can, apparently, be found if we distinct the directly detected, that is, changeable properties, such as relative mechanical momentum and energy, from the ascribed properties which are not detected directly; the last ones seem to be the persistent characteristics of the things in every event of their interaction however displayed by the directly detected properties. These persistent properties do not alter in a particular interaction process nevertheless they determine the specificity of influence and make the changes of the directly alterable properties to be specific. Such properties are thought to be inherent in things beyond interaction or detection. Inconsistency of this point of view was debated elsewhere (Esfeld & Lam 2005) because it is not clear how these allegedly non-relational properties become acknowledged through detectable, i.e. genuinely relational, properties. Detectable means influencing and causing change of another while concurrently influenced and changed by another. The non-relational properties can not be changed by their definition and can not contribute to interaction with others therefore they can not be detected, acknowledged, and measured. Nevertheless, we are aware of these allegedly non-relational properties and even enumerate them, i.e. we measure them by measuring reactions of things to external actions. If these properties are not non-relational, they must be classified as relational, but what to do with their persistence and changelessness? Indeed, some properties of things such as elementary electric charge or baryonic charge seem to be persistent and unchanging in any type of interaction, i.e. they look genuinely intrinsic, albeit mediatory acknowledged through the alterable relational properties. Perhaps, the predicate "in any type of interaction" is invalid? For example, mass is postulated unchanging in classical mechanics. However, it is linked to internal energy of things according to Einstein's formula $E = mc^2$ and this internal energy can be easily changed by a proper external influence; mass looses its nobleness in quantum mechanics. Unification of particles of opposite electrical charges, i.e. formation of ensembles of closely interacting things, gives birth to neutral objects for their environment (e.g. neutron consisting of one proton and one electron and existing as a superposition of wave functions of both, so it is impossible to identify where is the proton and where is the electron in the neutron). Annihilation of discrete matter, which is interaction of elementary particles with their antiparticles, leads to their disappearance for the world of discrete objects, that is, to alteration of all their charges. Color of macroscopic things depends on their internal structure and temperature and both can be altered in interaction with other things. It looks like there are the properties that persist within a class of actions while other properties are subjected to significant change. The first ones can be called intrinsic with the reference to this particular class of actions and others as non-intrinsic however none of them is non-relational. Mass is intrinsic in classical mechanics, but it looses the crown of intrinsicness in quantum mechanics. Elementary electric and baryonic charges can be considered intrinsic in virtually all interactions of discrete things except the processes of annihilation. Moreover, these elementary charges are the primary properties of discrete matter and they are directly related to physical vacuum that gives birth to all the charges of the discrete things on the microscopic scale of matter composition through interaction of the fundamental particles with the underlying vacuum (space) quanta. Here, the solution appears to be rooted. The fundamental charges of elementary particles appear through interaction of particles with underlying quantum vacuum as parameters that characterize the space tensions around each of the particles.  The space tensions dictate all the influences of things on each other and their reactions to the influences, and insofar as a specific charge is directly associated with a specifically tensioned space, we have to accept that this charge determines the specificity of changes of the alterable properties of others. If so, the persistent properties can not be called non-relational: their determinateness, displayed by the specificity of the directly detectable properties, is also acknowledged through the changes of others. It is quite interesting: these properties do not change within a class of interactions, but they are responsible for changes of alterable, i.e. genuinely relational properties. They characterize every influence of a thing on others (e.g. mechanical force of action) and, together with the 'intrinsic' properties of others, they determine every change of others as well as reaction (change) of this thing to action of others. Thus, it is consistent to assume that a signal sent by a discrete thing to other discrete things, that is, a distortion of space associated with corresponding charge, carries information about the 'intrinsic' properties of this influencing thing displayed by the particular change of the changeable properties of the influenced things. Every influenced thing bears information about the properties of the influencing things, including their 'intrinsic' properties, recorded in the change of its alterable properties. Intrinsic properties seem to be just conditionally intrinsic; these properties are not 'closed' inside a thing but reveal themselves to others by means of specificity of changes of directly alterable properties.

Virtually all the alterable properties of things can be reduced to their relative mechanical properties such as their relative position, momentum, kinetic energy, and potential energy. Color, rigidity, and many other properties of macroscopic objects as well as masses and internal energies of quantum mechanical micro-objects as wholes are determined by the structural composition of the corresponding ensembles, which, in its turn, depends on potential and kinetic energies of components. The structure of an ensemble is a particular spatial arrangement of the interacting mobile components established as a result of their interaction and characterized by their relative positions. Its specificity is determined, in particular, by potential energy of components in the fields of each other; the sum of potential energies of components yields internal potential energy of an ensemble and the sum of their relative kinetic energies yields its internal kinetic energy. Change of relative mechanical properties of a couple of interacting things results in change of their mutual influence on each other: every change of relative distance results in corresponding change of the force of mutual action and, therefore, in change of their relative potential energy which leads to alteration of internal potential energy of this couple as an ensemble. Change of their relative momentums results in change of their relative kinetic energy and, therefore, in change of internal kinetic energy of this couple. Change of total internal energy of an ensemble occurs only through interaction with the things external to this ensemble; every change of internal potential energy means some change of structure and, correspondingly, change of all properties of this ensemble dependent on the structure (color, transparency, rigidity, fluidity, form, and even mass of microscopic particles); every change of internal kinetic energy means change of relative motion of components and, correspondingly, change of all properties determined by this relative motion, such as temperature.

In a couple of interacting things, one thing influences another one and causes its change. Change of another means change of its original alterable properties and, therefore, change of its counterinfluence on the first thing leading to some conditional change of the first thing which bears now (in its change) information about the properties of the second thing imprinted in the alterable properties of the first thing in addition to its original properties. Change of the first thing means change of its influence on another one. The newly altered influence on another thing produces a new change of another and leads to further alteration of its properties and therefore to alteration of its influence on the first thing. It causes new change of the fist thing, which causes new change of another thing, and so on *ad infinitum*. Influence of a thing on others in a group of interacting things permanently alters in chord with the changing properties of all the things including this thing itself. The mutually conditional change means that every thing amid other things is not something absolutely stable, processless, and immobile; neither it is absolutely individual. Every thing is always in the mode of movement and transformation because of the changing influence of others subjected, in their turn, to the changing influence of this thing. The current change of an interacting thing occurs in accordance with the current properties of its counterparties while the current property of this thing is mirrored by the current changes of others. It can be said that a thing transfers its own property to other things at every particular moment of their interaction to be imprinted in their changes as if it tries to 'share' its own property with the counterparties and, concurrently, it accepts the shared properties of the counterparties imprinted in its own change. The outcome of this phenomenon is that the relevant properties, i.e. the properties displayed in a given type of interaction, become mutually shared to an extent. For example, velocity, spin, and trajectory of a billiard ball after a strike depend not only on its own properties but also on the cue's velocity, direction of strike, form of its tip, and conditions on the surface of the tip. The properties of a cue are recorded in the movement of the ball after the strike; the ball bears information about the cue. Conversely, the ball's properties such as elasticity, mass, and surface conditions determine the preferable properties of a cue to achieve a desired result. An outcome of collision of an ivory ball with a clay ball would be essentially dissimilar to its collision with another ivory ball. Every change of the influenced thing mirrors the properties of the influencing thing and vice versa; interaction is always the mutual sharing of the properties to an extent. Interchange with mechanical properties, such as momentums and kinetic energies, is an example of sharing of properties between the components in a system of mechanically interacting things. Interdependency of momentums of components relative to an arbitrary coordinate system results in appearance of a total (averaged) momentum of a whole system (ensemble). Permanent interchange with kinetic energies between the components gives rise to such a characteristic of an ensemble as its averaged kinetic energy per component, i.e. temperature, which is one of the emergent properties of every whole consisting of mechanically interacting components. Atoms in a molecule share their electrons from their highest electron shells and these electrons become common for all constituting atoms – atoms share their electromagnetic properties; in metals, 'gas' of shared electrons becomes common for all the atoms of a macroscopic object. Elementary particles in a nucleus share their nuclear properties to produce a sort of quantum superposition of their individual quantum states. In a neutron star, all the nuclei share their nuclear and electromagnetic properties to extreme indistinguishability and become a dynamic macroscopic quantum ensemble of neutrons. The phenomenon of sharing of properties is always dynamic in its essence because interaction means mutual influence by means of space distortions (physical fields) resulting in relative movement of interacting things. The outgrown properties of ensembles become relatively stable *on average* while the properties of their dynamic components are permanently, albeit conditionally, changing.

Information about the properties of another recorded in the changed property of a particular thing can be transferred to a third thing which comes to interaction with this thing simultaneously or even after its particular interaction with the first another. For example, a billiard ball reveals in its collision with another ball the changes of its mechanical parameters acquired from a cue. In the course or even after a particular event of interaction (by the event of interaction, we will understand a radical change of interacting things occurred during a finite time interval with much weaker interaction beyond this interval, so their mutual influence after the event can be neglected), each thing 'remembers' about its interacting partner and can present this memory to a third thing coming to interaction with it. This is the extremely important capacity of matter. Cognition of material things, which can not be directly observed or detected by our senses or tools, is possible solely because of such a transfer of information. Physicists study atoms and elementary particles making them to collide with each other like billiard balls to deduce their original properties from the outcome of their collision (proton structure, for example, was studied likewise). Astronomers find planets orbiting other stars without direct observation of the planets. They detect and measure the periodical velocity change of the stars caused by the gravitational pull of the planets to decipher, after measuring the star wobble, the properties of the planets such as mass, parameters of orbit, and distance between a planet and a star. Physicians diagnose deceases relying on medical tools that change their properties in interaction with our bodies. Geologists study minerals to understand processes of rock forging inside our planet billions years ago. Paleontologists study fossils to surmise appearance of prehistoric organisms and archeologists study artifacts to conjecture technology and life style of our predecessors. Our sensual perception is based, in fact, on such a mediatory transfer of information from one thing to another. We see a distant object because it either emits light or reflects light emitted by another source. Our eyes interact with the distorted electromagnetic waves and form images on the retina After processing the signals from the retina's sensitive cells, our mind ascribes to the objects their visible form and color. We hear sound without contacting a source of sound. The source oscillates and its oscillations force air (another physical thing) pressure to oscillate. The eardrums in our ears interact with air, being forced to oscillate by oscillating air pressure, and finally, the oscillating eardrums disturb the sensing receptors and we hear sound. Actually, we hear air and see light (photons), because our senses are impelled by them, however we have no doubt that we see the objects and hear the sources of sound. All our communication devices employ an altering thing as the source of a signal about its changing properties recorded in the change of another thing or medium (a carrier). A detector (or a receptor of our senses), also interacting with the carrier, alters when it receives the signal about the changed properties of the carrier. We observe the detector's change by means of our senses, meaning that our senses (receptors) react to the changing detector and alter, too. Eventually, our nerves, influenced by the changing receptors, convey the signal about the receptors' change to our brain. Often several carriers consecutively participate in transmitting the signal: space, optical materials, semiconductors, metallic wires, air, etc.

**3.2 Togetherness of interacting things**

Influence and counterinfluence of interacting individuals culminate in their mutually dependent changes. Interdependency of changes means that all parts are united in an ensemble of components with mutually shared and interdependent properties (ensemble, meaning 'together', seems to be suitable for describing the phenomenon of togetherness of interacting individuals). Through interaction with other components, each component 'lays claim' to its own existence as a participating partner in a whole. A particular individual is united with other components in a kind of collective unity in which all the components share their properties with each other and bear the properties of each other in their changes. An ensemble of things appears as a new thing with its own properties emerging through the process of sharing of properties between all the components. Its existence supervenes upon existence of its own components however this supervenience is not a direct sequel of existence of separate components with their pristine individual properties but a consequence of their together-being. Every ensemble attains its properties through interdependency of the distortions of the properties of all the components in the process of dynamic averaging of their changes over a period of time needed for sharing properties between the components. As existing something and, therefore, as a bearer of its own property, every ensemble becomes a thing for other things around it, i.e. a physical object amid other physical objects. Atoms (ensembles of nuclei and electrons) possess their properties that differ from the properties of nuclei and electrons taken separately, molecules (ensembles of interacting atoms) essentially differ from their own individual atoms, water (an ensemble of molecules consisting of hydrogen and oxygen atoms) is a different substance in comparison with hydrogen and oxygen gases, and a snowflake (an ensemble of closely and strongly interacting water molecules) is very dissimilar to a cloud of vapor (an ensemble of loosely interacting water molecules) despite the fact that both consist of the same molecular species. The properties of ensembles seem to be alien to the properties of their own components. For example, conductivity and superconductivity of substances are not linked directly to the properties of atoms or molecules composing them; fluidity and superfluidity (at low temperatures) as well as temperature of macroscopic objects can not be attributed to individual atoms or molecules; crystalline structure and electrophysical properties of macroscopic objects can not be expressed in terms of properties of individual molecules; luster of metals and color of objects in general have nothing in common with absorption or emission spectra of constituent molecules or atoms, spiral structure of galaxies can not be explained in terms of separate and independent stars orbiting a galactic center, etc. There is no known successful attempt to derive the second law of thermodynamics from the equations of classical mechanics without some additional assumptions actually introducing chaos and irreversibility.(Tisza 1966) New properties of ensembles as physical objects appear through the collective being of the constituting individuals; their alienation from the properties of their components gives us a right to presume that the properties of all the material objects as ensembles are ontologically emergent. A collective Many of interdependent individuals can not be simply reduced to a single constituent or to a partition; the collective properties of a Many, perceived as a One, seem to ontologically emerge.

The maximally dramatic change occurs to those thing's properties which undergo the fastest alteration: the stronger is influence, the faster and more profound is change, and the stronger are bonds that bound the interacting counterparties. Change of components in an ensemble occurs faster and deeper through interaction with each other than because of their interaction with the things external to this ensemble because their mutual co-influence is normally stronger than any influence originated from external (remote) things. For example, a billiard ball (structural ensemble of strongly bounded molecules) keep its form and internal structure after a strike (after an event of interaction of the ball, as an ensemble of molecules, with the cue which is another ensemble external to the ball) despite a noticeable change of its state of movement with respect to the cue and other balls. Interaction of molecules inside the ball is much more intense than their interaction with an external macroscopic object under normal conditions. Intermolecular ties inside the ball provide the ball's integrity and keep its form and structure virtually intact. Interaction of components in an ensemble is virtually always stronger than interaction of this ensemble with other ensembles in a superensemble. Structure of atomic nuclei stays virtually unchanged in the processes of excitation, ionization, or recombinaion of atoms. Internal structure of atoms do not noticeably alter when a molecular substance appears or a solid body forms from a mixture of free atoms despite the fact that other attributes, such as relative positions of atoms, their momentums, and even form and structure of external electron orbits in atoms undergo significant changes. In general, deepness of the mutual changes of interacting things and rapidity of these changes are determined by severity of their influence on each other. Severity of influence can be characterized by a 'force of action' which stems from asymmetry of tension of space around each of interacting things: the higher is asymmetry, the larger is disparity of space tensions at the opposite edges of a particular thing and, therefore, the stronger is mechanical force acting on this thing. The deepness of change of an influenced thing is proportional to the force of action and a time interval of action; the stronger is the force of action, the faster is accumulation of changes during a certain time interval and the deeper is the total change at the end this interval. If, by any reason, an external force of action on a particular thing is stronger than the bounding ties between its components, this thing looses its integrity. Colliding elementary particles behave like billiard balls unless their relative collision velocity (kinetic energy) is sufficient to overcome their electrostatic repulsion, to overcome the ties of quarks in them, and to cause nuclear reactions; molecules can be considered as solid balls in collisions with each other unless the force of mutual action becomes sufficient for their dissociation into atoms, for example, dissociation of gas molecules at high temperatures; a billiard ball stricken by a bullet will loose its wholeness, etc. As a rule, the ensembles, which are closer to the level of fundamental bricks of matter (quarks), can keep their integrity intact in the processes of interaction of ensembles on the higher levels of matter composition. The structure of nuclei and elementary particles stays unaltered in chemical transformations of molecules; mechanical interaction of macroscopic objects does not change noticeably their molecular structure; remote gravitational action of celestial objects exerts no noticeable influence on macroscopic objects on the surface of our planet, etc. Interactions characterized by a stronger force and a higher rapidity of changes take place at a lower structural level of matter. Ensembles formed on a lower structural level of matter can become components of an ensemble of a higher structural level; the ensembles that belong to this higher level form ensembles on an even higher structural level (superensemble), and so on. Our world is a hierarchy of ensembles where every physical body is a whole of interacting composing things (subensembles) and, concurrently, is a part of a superensemble of interacting ensembles. A number of varieties of ensembles increases dramatically with elevation from the lowest structural level of matter. Four types of quarks produce a family of elementary particles; only a few elementary particles (the proton, the neutron, and the electron) [5] form more than a hundred of chemical elements (varieties of atomic ensembles); chemical elements give birth to thousands and thousands of molecular species; molecules, in their turn, form innumerable varieties of substances and physical bodies. Multiplicity of mutual bonds and, therefore, multiplicity of arrangements of material entities increases progressively with a growing number of interacting components of different species. Specificity of interaction of things plays a major role in forging the ensembles and in chiseling their appearance for others, i.e. their quality.

It is important to mention that in a mishmash of asymmetric distortions of space (spatial distributions of fields) around multiple discrete things there are only two ways for building a relatively stable spatial structure: either repulsion is counterbalanced by attraction or all the interacting things possess some relative motion. In the last case, relative mechanical momentums of interacting things can balance their mutual attraction/repulsion and cause their oscillatory (periodical) movement, for example, motion of electrons in atoms, oscillations of molecules near the points of equilibrium in solids, revolution of planets around a star, etc. It is one of the most remarkable features of matter: movement, associated with alteration, inconstancy, and instability, and wholeness, associated with stability and persistence, are paradoxically linked to each other. Wholeness arises from interdependency of changes of relatively mobile components in their together-being. It results from changeability and outgrows from inconstancy. There would be no whole without the mutual co-change of parts. Interdependency of the modes of being of interacting things results in their collective being with mutually restricted relative motion. Due to this restriction, something more structural, more stable, and more persistent appears than the parts permanently changing each other. The constraint, imposed by one thing upon another, together with the reciprocal constraint give birth to a unity of mutually constrained (therefore, bounded together) things with their coordinated movement. An ensemble of things emerges, which keeps its bounded state and, therefore, its form, structure, color, etc. during all the period of mutual constrain. The lifetime of an ensemble can be incommensurably longer than the time interval of passage of otherwise non-interactive things near each other or the collision time during which these things radically change their relative movement, for example, their period of oscillation, period of orbital motion, average interval between two consecutive collisions such as collisions of molecules in gases, etc. Swarming molecules can form a relatively stable macroscopic thing when their mutual electromagnetic bonds are sufficiently strong. Molecules do not lose their relative movement but transform it to a sort of constrained motion, such as oscillatory motion in solids or chaotic motion between collisions in gases. The lifetimes of binary stars and planetary systems are much longer than the time intervals of passage of celestial objects near a host star without capture. Stars in a galaxy are in dynamic constraint imposed by gravitation of the totality of stars inhabiting this galaxy; their togetherness results in formation of a structural galactic disk with its lifetime exceeding the lifetimes of the stars as shining objects. Stability of wholes outgrows from dynamism of components constrained by their mutual interaction.

No object in the material world is genuinely individual in a sense that it is not an ensemble of interacting components. Quarks are thought to be simple according to the Standard Model, however they are unable to exist as separate individuals: only ensembles of quarks, i.e. elementary particles, are real and possess locality.(Griffiths 1987a) Every component of an ensemble undergoes changes induced by all other components from various transient distances. Since the 'force of action' is the perceived action of asymmetrically tensioned space created by each of the components at a location of another component, it depends on the distance between them. The components located farther from a particular component exert weaker direct influence on it. As a part of an ensemble, every component is bathing in the distortions of space (in physical fields) spawned by all other components. Their mutual influence, either direct or mediatory, i.e. through other components, and their mutual sharing of their mechanical properties lead to such a balanced disturbance of space around each component and to such their relative movement that, on average, they keep their relative distances from each other approximately (on average) constant – a spatial structure emerges as the way of collective being of mobile components in their togetherness.

All things in the Universe as a superensemble of ensembles have interacted since the moment of becoming and, to lesser or larger extent, they will always interact (at least, within the event horizon). The objects separated farther from each other exhibit relatively weaker influence on each other and their mutual changes are slower. In addition, the limited velocity of propagation of changes of space distortions in space (speed of light) imposes some temporal restraint on the rapidity of their interchange with properties: the farther are the objects from each other, the longer is the time interval between the consecutive exchanges with the properties and the larger is the time interval for establishing a structure and to form a stable ensemble. Influence of extremely remote objects (signals of their change), e.g. gravitational action of other galaxies on our planet at current epoch of Universe's evolution, was originated from them millions or billions years ago. A time scale for establishing wholeness of ensembles is different on different levels of matter composition. A whole emerges as soon as all the components interchange with their properties and constrain each other's relative motion.

Interdependency of things in the Universe was formulated originally as Mach's principle. Mach argued that such a property as inertia is not innate in a physical body, but arises from its relation to other remote objects. The remote others establish a local inertial frame by their mere existence and determine inertia of a particular thing in contradiction with the idea of absolute intrinsicness of this property. "When, accordingly, we say that a body preserves unchanged its direction and velocity in space, our assertion is nothing more or less than an abbreviated reference to the entire universe" (Mach 1960). The Mach's idea was ingenious. For the first time it was guessed that the mode of being of one thing depends on the totality of things in the Universe. This idea was generalized by Einstein (see, for example, (Stachel 1989)). According to his theory of general relativity, the collective gravitational action of all things in the Universe produces an overall curvature of space and determines its metrics; this metrics of space controls the mechanical behavior of every particular material thing at every location. In particular, the mentioned flatness of our space (on average) and applicability of Euclidian geometry is the outcome of approximate equality of matter density (conventional mass + vacuum energy) to critical density. Metrics of space is completely determined by the distribution of mass-energy in the Universe, i.e. by the totality of gravitationally interacting objects, and every object obeys this metrics and displays its observed mechanical properties (Barbour & Pfister 1995). "The best interpretation of our best science tells us that the properties of things in the world may not be fixed absolutely with respect to some unchanging space-time background, but rather that these properties arise from their interactions with and relationship among the other things in the world." (Silverstein & McGeever 1999)

Since the distortions of the properties bear information about the properties of each of the influencing things, every component of an ensemble bear information about all other components and therefore about the whole ensemble. It means that the accepted (shared) enumerated property of a particular component characterizes the whole ensemble because every particular change of a component reflects the feedback reactions of other components and vice versa. In a balanced state, a sort of dynamic equilibrium is reached, when, on average, [6] all involved in mutual interaction components and, therefore, the emergent properties of this ensemble do not change any more. An ensemble itself becomes relatively stable, albeit internally dynamic, with continuously, albeit interdependently, changing components. The components behave themselves as if they are controlled by this whole ensemble which 'forces' them to be disciplined. It can be said that every ensemble influences its own components and transfers its properties to them as any other influencing thing; this phenomenon gave birth to the concept of inverse or downward causation.(Andersen et al. 2000, Campbell 1974)

Thus, observing a portion of a whole or even one component during a sufficiently long period of time, one can judge about the properties of this whole. Studying a small portion of a crystal, we judge about total structure of this crystal. Studying a sample of a substance, we judge about the properties of this substance. To measure a temperature of an object, we put our thermometer in contact with a small area on its surface, i.e. in contact with a small part of a whole; nevertheless, we confidently assess the temperature of this whole object. In the process of measurements (interaction of a tool and an object), our thermometer and a measured object are united in a superensemble; both become interrelated and their temperatures equalize. Finally, we get a reading of temperature of this ensemble, which can be taken as the temperature of the measured object on the condition that our thermometer produces relatively small impact on their common temperature or if the object is thermostatic.

Interaction of physical objects, in particular their mechanical interaction, is usually regarded as their direct action on each other as wholes. It is important to emphasize that this conviction is totally wrong. The components of ensembles are the entities that produce influence on all other things including the external-to-ensemble things because they are the sources of spatial distortions (physical fields). They produce influence on all others and they are subjected to change under influence of others, be these others the components of this ensemble or things external to it. Reciprocally, all the external things act on the components of a particular ensemble but not on the ensemble itself despite its apparent reaction as a whole. A change of externally influenced components (not necessary all the components: a portion of them can be influenced or even one component only) is transferred to all other components of this ensemble through their mutual bonds. It results in a corresponding reaction of all components and produces the observed response of an ensemble perceived as its change as a whole. A billiard ball, stricken at a small area on its surface where only a small portion of molecules is directly influenced by a portion of molecules of a cue, reacts as a whole because the influenced molecules transfer their mechanical momentums acquired from the cue to all other molecules of this ball through their elastic bonds keeping them together in a structural arrangement. Interaction (collision) of a molecule with another molecule is nothing else but interaction of their atoms at a point of maximum proximity to each other, that is, interaction of a portion of one ensemble with a portion of another one. If the force of interaction of the components of one ensemble with the components of another ensemble is less than the strength of their bonds, in other words, if this force is not sufficient to disparate the components from their ensembles, both ensembles react as wholes and change their properties as wholes.

The prime bricks of matter (quarks) seem to possess the unique ability of generating distortions of space (fields according to the field concept) and, correspondingly, to react to all external distortions of space (fields). They are responsible for nuclear, electrostatic, and gravitational interactions of all ensembles on all levels of matter composition. Concurrently, they are the primal entities that respond to all actions of things external to any ensemble on any structural level. Every thing external to a particular ensemble produces action primarily on quarks and electrons; quarks, confined by much stronger nuclear ties, transfer their reaction and change to ensembles of quarks, i.e. to elementary particles, and cause their mediatory response; elementary particles, tied in nuclei by strong nuclear interaction, transfer their change to nuclei as ensembles of elementary particles; nuclei and electrons, tied in atoms by electromagnetic interaction which is normally stronger than an external influence, transfer their acquired change to atoms and cause their acceleration, excitation, or polarization; eventually atoms, tied together in molecules and in macroscopic objects by electromagnetic or gravitational forces transfer their acquired change to physical bodies. Every response of an ensemble outgrows from the responses of its components down to the prime material modicums ultimately causing all responses of all material things. Every interaction of one ensemble with another ensemble results in change of both, but it should not be forgotten that the components of one ensemble (often only a few of them) are directly influenced by the components (not necessary all of them) of another ensemble. The observed responses of things as wholes are always mediatory and outgrow from the responses of their own components bound together by their mutual ties.

Now, after the detailed consideration of the phenomenon of togetherness of things resulting in formation of material wholes (ensembles), it is appropriate to discuss the allegedly uncaused processes such as radioactive decay. Actually, radioactive decay is a particular case of significantly more general phenomenon of spontaneous radiation. It must be emphasized that all the processes of spontaneous radiation occur in ensembles, more precisely, in quantum ensembles. All ensembles consist of mobile, interacting, and mutually influencing/influenced components bounded by forces of their mutual action. A particular state of an ensemble (its structural variety) can be characterized by a value that enumerates the potential energy of all its components in the fields of their mutual bonds, that is, in the distortions of space spawned by each of them (some aspects of potential energy of quantum ensembles were discussed in (Teller 1986)). The enumerated potential energy of one thing in the field (distortion of space) generated by another thing is determined as convolution of acting force and distance between the things. Potential energy of a pair of things increases when the distance between them increases. When the distance decreases, potential energy of this pair diminishes and its subtracted part either transforms to kinetic energy of relative movement of things or emitted outward to their environment in the form of radiation (waves of space tensions or fields). The notion of potential energy is only meaningful for a system of interacting things and, therefore, it characterizes every ensemble of interacting things. Contrary to classical mechanics of macro-world, enumerated potential energy of quantum ensembles can only accept some discrete values. The quantum ensembles change their state and internal structure jumping from one level of internal potential energy to another level. Spontaneous radiation can occur if, and only if, a quantum ensemble had been previously transferred to a state characterized by some higher level of potential energy of internal bonds of its components, i.e. from a relatively stable state on a lower level of potential energy (as a rule, from the lowest level corresponding to the highest degree of bondage of components and the highest stability) to a non-equilibrium and unstable state on a higher level of potential energy with a lower bondage of components. The process of excitation is only possible if the excess of energy for the jump is taken from outside, e.g. from potential energy of another ensemble, from kinetic energy of relative motion of another ensemble when it comes to interaction, or from energy of oscillating space tensions (waves of fields). If such an excited ensemble jumps afterwards back to a lower potential energy level because of the inherent instability of its state on a higher level, the earlier captured potential energy is released outward as spontaneous radiation (outward means to space and finally to the external things). Radioactive elements on our planet together with all elements of Mendeleev's table heavier than lithium were forged in the interiors of stars and in the explosions of novae and supernovae existed even before our Sun was formed with its planetary system (Clayton 1983) (some radioactivity can be produced by cosmic rays, too). Radioactive elements became radioactive just because their nuclei as quantum ensembles of elementary particles had been excited in their past on a higher level of their internal potential energy in collisions with other high energy particles. Radioactive decay is transition of the previously excited nucleus from a higher level of potential energy to a lower level. Likewise, spontaneous electromagnetic radiation of atoms occurs after excitation of these atoms to some higher energy levels in collision with electrons (e.g., luminescent lamps in your apartment or office), with other atoms, or due to absorption of a photon emitted by another atom. In all cases, transition of a quantum ensemble to a higher level of potential energy is caused by external action. The real question is: what is the cause of the backward transition which appears as happening by itself, i.e. without external influence, and why the excited ensembles tend to return back to their unexcited state? Detailed dialectics of non-equilibrium ensembles is beyond the objectives of this article; it is sufficient to mention here that quantum ensembles consist of interacting mobile components and that every internally dynamic ensemble tends to its dynamic equilibrium state characterized by the lowest internal potential energy. Energy, released in transition from a higher level to a lower level of potential energy, has to be transferred to something material, because energy in general is a characteristic of matter, not a void. A hypothetical isolated quantum ensemble would exist forever in an excited state (mathematically, each quantum state is the exact solution and such an ensemble has no reason to jump anywhere 'by itself'). In reality, there are always other discrete things permanently influencing a particular thing and disturbing its pristine quantum state; moreover, there is a material substance which is always in interaction with every discrete thing: space. To be in interaction means to influence and to be influenced, and interaction of discrete matter and space is not an exception. From the point of view of quantum field theory, spontaneous radiation, i.e. the transition of a quantum ensemble from a higher level of internal potential energy to some lower level, is caused directly by the zero oscillations of vacuum fields, if all the direct influences of other discrete things are excluded. In particular, the virtual electromagnetic field of the oscillated virtual electron/positron pairs in physical vacuum (space) pushes the exited electrons of atoms from their higher energy levels to produce 'spontaneous' electromagnetic emission; likewise, radioactive decay of nuclei is directly caused by zero oscillations of vacuum nuclear fields. "…There is no such thing as truly spontaneous emission; its all stimulated emission".(Griffiths 1987b) Energy, released in transition of an ensemble to a lower level of internal potential energy, goes to the waves of fields in space (waves of tensions of space), e.g. to electromagnetic waves such as photons and gamma-quanta, to gravitational waves, or to nuclear waves in the form of bosons.

Transition of a quantum ensemble from a higher to a lower energy level causes change of this ensemble, which, in its turn, causes change of space displayed by the waves of tension of space propagating outward with a finite velocity (speed of light); this wave can be detected by another discrete thing (detector) reacting to the change of space tension around it when the wave reaches the detector. Observing the change (reading) of our detector, we become aware of the change of the thing and can measure this change. However, we have to interpret correctly the detector's change, which addresses us to correctness of the methods of measurements including our choice of mathematical procedures. (Puri 1985)

## 4. CONCLUSION

Interaction is a key category for understanding the phenomenon of wholeness in the material world; it is actuality that lies in the very foundation of our Universe. The mode of being of a material thing is its interaction with other things. Through their mutual interaction and restrain, interacting things form a sort of collective being, i.e. an ensemble of things with coordinated motion of components and with their interdependent properties. All material objects in the world are ensembles which come to being through togetherness of their components. Their wholeness means that every ensemble is composed of parts with interdependent modes of being and that the mode of being of one part depends necessarily on the modes of being of other parts of this ensemble and *vice versa*. Every ensemble attains its structural wholeness and realizes its own existence as an unabridged whole through correlation and mutual 'cooperation' of individual components. Its existence and locality is a result of continuous interaction of mutually changing and constraining each other parts. It is the process of becoming at every particular instance from the originally free and relatively mobile composing things after they constrain each other through their mutual influence to become entangled in the collective being. The collective mode of being and coordination means that formation of ensembles as material wholes is to be understood as ontological emergence because there are the properties of ensembles which can not be attributed to their individual parts.

Emergence of properties of ensembles is not an exclusive prerogative of quantum mechanics. Togetherness of interacting things, their correlative behavior, and their entanglement give birth to emergent properties characterizing every material thing on every structural level of matter composition. "Emergent properties are properties of a system taken as a whole which exert a casual influence on the parts of the system consistent with, but distinct from, the casual capacities of the part themselves."(Silverstein & McGeever 1999) The phenomenon of reverse causality conceived as the apparent 'reverse' action of a whole on its own parts is equally applicable to microscopic (quantum mechanical) and macroscopic objects. However, it should not be forgotten that the components with their interactions are the prime players that impose coordinated behavior on themselves through their mutual influence, which is perceived as a property of this whole and as its influence on its own parts. The confusion arises, perhaps, from the conviction that emergency and irreducibility somehow negate causality. On the contrary, every ontological emergence is causal. Collective behavior lies in co-relations; co-relations are to be understood as mutual influence and as interdependent change of interacting things which leads to the phenomenon of sharing of the properties with each other. From this point of view, 'relation' already implies correlation and, therefore, togetherness and interdependence of properties. A single thing, i.e. a thing without others, is absurd. Relations and properties are nonsensical without others. The properties of a lone thing, if we assume any, can not be co-related and acknowledged, which is equal to nonexistence. Everything seems to lie in relations; the whole world is fashioned by them. "Objects are not atoms that exist independently of each other and… structure always consists of specific, concrete relations, this relations being as determinate as intrinsic properties are supposed to be."(Esfeld & Lam 2005) The being of a thing as a monistic entity is its being as an ensemble, i.e. as a collective being of a number of parts which are firstly mobile and secondly influencing and constraining each other during all the period of their together-being. Correlation of parts culminates in giving birth at each particular moment to a finite structural entity with its ontologically emergent properties. Existence of all things in the world is their continuous emergence through maintaining the coordinated behavior of their components. It is not a single event of establishing of properties at some instant forever after; it is continuous and uninterrupted genesis of properties of physical objects through the maintained togetherness of parts. Emergence is process (Ahearn 1994) which is not simple motion but genesis. The important feature of this genesis is that the components of an ensemble become entangled and impart a quality to this ensemble during a characteristic time interval needed for sharing the properties between all the spatially separated parts.

The vision of the world as a hierarchy of ensembles of things with their properties continuously emergent from the collective beings of their components, which, in their turn, are the ensembles on a lower level of matter composition, gives us confidence to believe that relational holism reflects a real feature of the whole material world. "In such a world there will be no discrete hierarchy of casually closed levels."(Silverstein & McGeever 1999) Interdependency of properties of components in a particular whole implies a sort of inseparability: take off a part of a whole and you obtain, generally speaking, another whole, because the newly obtained whole is not a result of simple numerical subtraction but is to be understood as the absence of affect of the excluded part on the collective being of the rest of the parts. The excluded part was the necessary and inseparable portion for establishing the properties of the former whole.

If every thing is a part of a hierarchy of ensembles, what is considered to be true for the underlying microscopic must also be true for the emergent macroscopic. Every thing is ultimately composed of quantum mechanical entities and their features are inevitably transferred to all higher levels of matter composition, because it is the microscopic fundamental particles that are primarily subjected to influence and act on others in all interactions of all ensembles. It can be shown mathematically that every ensemble of macroscopic components, subjected to a very small but finite chaotization of momentums in collision (interaction) with each other due to quantum mechanical nature and inherent probabilistic (chaotic) behavior of their own microscopic constituents, inevitably evolves to Gaussian (chaotic) distribution of momentums of these macroscopic components in the ensemble, i.e. to its equilibrium state, provided the time interval of observation is sufficiently long. "First, ontological emergence within quantum mechanics makes it plausible that it exist elsewhere, even if it is not quantum mechanical in nature. Second, either everything is reducible to fundamental physics or it is not. If it is reducible, if everything is quantum mechanical, then ontological emergence is ubiquitous... If, on the other hand, the macroscopic is not reducible to the microscopic, if quantum effects are really screened off, then the entire macroscopic world is somehow ontologically emergent. In short, ontological emergence is most probably a real feature of the world. Ontological emergence means monism without reductionism."(Silverstein & McGeever 1999)

The supposedly 'intrinsic' properties of physical objects, such as color, temperature, and even mass, are outgrown from specificity of interaction of components in their collective being and can be changed by proper external influences. Perhaps, the properties of prime fundamental bricks of matter such as their electric charge, baryonic charge, charm, etc. can be assumed intrinsic if we forget about physical vacuum (about materiality of space) in which the origin of these apparently intrinsic properties of fundamental particles is rooted according to quantum field theory. A tentative boundary between supposedly intrinsic and extrinsic properties of things can be found, if the properties that are not directly changed by external influence under normal conditions and within a certain class of interactions are considered as intrinsic while other properties subjected to direct change as non-intrinsic; none of them, however, is non-relational. Though the supposedly intrinsic properties do not change within a class of interactions, they actually determine specificity of changes of the alterable properties and become acknowledgeable and mediatory measurable through them. Absolutely 'intrinsic' or closed properties are abstractions related to isolated and therefore ideal things. Real things are never isolated. They are together. Their properties are variable. Their change is absolute.

Interaction is the way of being of a being amid other beings. To be is to interact – to manifest and to be manifested, to meet others and to be met by them, to influence and to be influenced, to change and to be changed, to share and to be shared, to unite and to be united – to be together. Togetherness of individuals and wholeness in general fashion the material world and drive it from variability to stability, from separateness to consolidation, from Many to One, and from dull alikeness to beauty of structural diversity.

**NOTES**

[1] According to quantum electrodynamics, physical vacuum, which we perceive as space, can be considered as consisting of coupled (annihilated) matter and antimatter pairs on the lowest quantum energy level (Griffiths 1987a and b, Moriyasu 1983). In particular, the coupled pairs of the electron and the positron constitute so-called Dirac vacuum. Oscillating pairs generate virtual photons in space (spontaneously appearing and disappearing bubbles of electromagnetic field due to occasional small separations of the coupled particles) and these virtual photons (also called vacuum electromagnetic oscillations) are observed experimentally through their influence on conventional matter. For example, they cause Lamb shift of spectral lines in atoms and produce mechanical action on macroscopic objects (Casimir force). Thus, physical vacuum and, therefore, space is able to respond to the discrete electric charges through responses of the annihilated pairs to existence of the charges. This ability of space to respond to electrically charged objects is displayed by its finite permittivity $\varepsilon_0$ and permeability $\mu_0$. Permittivity and permeability characterize electromagnetic 'elasticity' of space and determine the speed of propagation of electromagnetic waves through space $C = 1/(\varepsilon_0\mu_0)^{1/2}$, i.e. propagation of changes of electromagnetic polarization of space.

[2] We live in three-dimensional space, and its three-dimensionality seems to be closely linked with the fact of existence of physical objects (discrete matter) in their observed forms. A two-dimensional (2D) world would be meaningless: no material particle can exist in it. In 2D-space, the condition of constancy of total force of space tension (total flux of field's strength according to field theory) acting on a spherical surface enveloping a particular thing means that the 'strength of tension' and the force of action of one thing on another would diminish as $1/R$. It leads to logarithmic divergence of tensile energy (potential energy of interaction of things) contained in space around a thing when $R \rightarrow \infty$. Therefore, infinite energy would be needed for a single thing to come to existence. In its turn, the tensile energy of space and the force of action between things in 4D-space would diminish as $1/R^3$ resulting in much shorter distances of action. 4D-space is inconsistent with quantum electrodynamics because no stable electron shell in atoms can exist due to significantly smaller electron orbits in comparison with 3D-space and high probability of electron capture by a nucleus (K-capture). No stable atomic structure can exist in 4D-space.

[3] Non-interactability of things and, therefore, their nonexistence resulting in their unobservability *in principal* should not be mixed with unobservability *ad hoc*, i.e. with inability of our senses or tools,

currently at hand, to detect something despite the principal ability of this something to interact and its potentiality to be detected by the advanced detectors (for example, objects in quest like the new celestial bodies, new elementary particles, dark matter, etc.) Actually, every search of a new object is nothing else but the hunt for traces of influence imposed on the already known objects or on the detectors. Hypothetic holes (Casati & Varzi 1994) in the material reality is to be associated with voids not in a sense of trivial absence of conventional matter in space but rather with the absence of space together with its metric. Such immaterial entity can not influence material things and can not constrain the modes of being of material things therefore it lacks properties for perceiving and detecting and therefore it can not be real.

[4] The counterexamples could be eavesdropping or observations of celestial objects which seem to occur without reciprocal action. Actually, eavesdropping is simply undetected presence (and influence) and can be discovered by a cautious individual or an anti-bug device. As for the celestial objects, we act on them gravitationally and electromagnetically concurrently with their action on us; however, our reciprocal action is miniscule to be observable and postponed in time because of the vast distances to these objects and the finite speed of propagation of signals through space.

[5] Quantum field theory, known as the Standard Model of elementary particles, unifies three types of interaction: nuclear strong interaction, weak interaction, and electromagnetism. Ordinary matter is considered as constructed of the electron, the up quark, and the down quark. Triplets of quarks bind together by gluons to form protons and neutrons, interaction of which, in their turn, makes up atomic nuclei. The electron, the up and down quarks together with the electron neutrino form the first of three groups of particles, called generations. Other two groups contain heavier and unstable analogs of particles of the first group. The spin-1 bosons are represented by the photon (carrier of electromagnetic interaction), the gluon (carrier of strong interaction), and the W and Z bosons (carriers of week interaction). Together with the spin-0 Higgs boson (hypothetical quantum of Higgs vacuum field), they compile a family of fundamental particles comprising 17 entities (nine 'bricks' of matter with three neutrinos and five carriers of force playing the role of 'mortar'). The model is fairly elegant and consistent with all known experimental physics up to currently reachable energies of accelerators.

[6] The properties of ensembles emerge from the collective beings of their components with mutually shared properties. The components are always spatially separated and have finite relative velocities. Every change of a state of movement of a component in interaction with another component is durable processes because a finite time interval is needed to cover the distance between the components and to 'collide', i.e. to undergo a radical change comparable with the absolute value of the particular

enumerated property. Ratio of the average distance between the components and their average velocity determines the average time between their 'collisions', i.e. the characteristic time of interaction which establishes the time scale for sharing (averaging) of properties between all the components. Thermodynamic properties of ensembles, such as temperature, are outgrown from the chaotic relative movement of the components and only meaningful if they are measured on a time interval exceeding the characteristic collision time or, alternatively, if they are averaged over a statistically large number of components at a particular instant. Multiple changes of dynamic properties of components due to their simultaneous interaction with multiple other components lead to uncertainty of determination of their exact dynamic properties at a particular instant. Every determination of exact parameters of a component requires knowledge of exact positions and momentums of all components at all previous moments. Even in classical mechanics, the multi-body problem is practically unsolvable. The notorious probabilistic behavior of quantum entities makes it unsolvable *in principle*. The solution is to describe the dynamic properties of the ensembles consisting of multiple components statistically. A particular enumerated property of an ensemble is described *on average*. We do not prescribe the exact dynamic parameter to every component at every moment, but rather calculate the probabilities to have this or that value with subsequent integration (mathematical weighing) to obtain the averaged enumerated dynamic property of this ensemble or the corresponding averaged measure per component (e.g. temperature which is averaged kinetic energy per molecule). This averaged enumerated property per component is not the characteristic of a particular component but the property of a whole ensemble because all the components of this ensemble with all varieties of their dynamic parameters contribute to this averaged value.